\documentclass[acmsmall,screen]{acmart}
 \acmConference[Preprint]{Preprint}{Preprint}

\usepackage{algorithmic}
\usepackage{graphicx}
\usepackage{textcomp}
\usepackage{xcolor}
\usepackage{subcaption}
\usepackage{pifont}
\usepackage{fancyhdr}
\usepackage{tcolorbox}
\usepackage{alphabeta}
\usepackage{multirow}
\usepackage{multicol}
\usepackage{booktabs}
\usepackage{wrapfig}
\usepackage{listings}
\usepackage{xcolor}

\begin{document}

\title{An Empirical Study of Vulnerability Detection using Federated Learning 
}

\author{Peiheng Zhou}
\affiliation{%
  \institution{East China Normal University}
  \city{Shanghai}
  \country{China}
}

\author{Ming Hu}
\authornote{Corresponding author.}
\affiliation{%
  \institution{Singapore Management University}
  \city{Singapore}
  \country{Singapore}
}
\email{hu.ming@ntu.edu.sg}

\author{Xingrun Quan}
\affiliation{%
  \institution{East China Normal University}
  \city{Shanghai}
  \country{China}
}

\author{Yawen Peng}
\affiliation{%
  \institution{East China Normal University}
  \city{Shanghai}
  \country{China}
}

\author{Xiaofei Xie}
\affiliation{%
  \institution{Singapore Management University}
  \city{Singapore}
  \country{Singapore}
}

\author{Yanxin Yang}
\affiliation{%
  \institution{East China Normal University}
  \city{Shanghai}
  \country{China}
}

\author{Chengwei Liu}
\affiliation{%
  \institution{Nanyang Technological University}
  \city{Singapore}
  \country{Singapore}
}

\author{Yueming Wu}
\affiliation{%
  \institution{Nanyang Technological University}
  \city{Singapore}
  \country{Singapore}
}

\author{Mingsong Chen*}
\affiliation{%
  \institution{East China Normal University}
  \city{Shanghai}
  \country{China}
}
\email{mschen@sei.ecnu.edu.cn}

\begin{abstract}



Although Deep Learning (DL) techniques, especially Large Language Models (LLMs), are becoming increasingly popular in vulnerability detection, 
their performance is seriously limited by insufficient training data.
This is mainly because few existing software companies or organizations can maintain a complete set of high-quality samples for DL-based vulnerability detection. Due to the concerns about privacy leakage, most of them are reluctant to share such samples, thus resulting in the {\it data silo problem}. 
%
As an emerging distributed DL paradigm, Federated Learning (FL) enables model training among clients without compromising their data privacy. 
Therefore, it has been investigated as a promising means of addressing the data silo problem in DL-based vulnerability detection. 
However, since existing FL-based vulnerability detection methods focus on specific applications, 
it is still far unclear i) {\it how well FL adapts to common vulnerability detection tasks} and ii) {\it how to design a high-performance FL solution for a specific vulnerability detection task}.
To answer these two questions, this paper first proposes VulFL,  
an effective evaluation framework for FL-based vulnerability detection.
%
Then, based on VulFL, this paper conducts a comprehensive study on 
a well-known real-world dataset named DiverseVul, aiming to reveal the 
underlying capabilities of FL in dealing with different types of Common Weakness Enumeration (CWEs), especially when facing various data heterogeneity scenarios. 
%
Our experimental results show that, 
compared to independent training, 
FL can significantly improve the detection performance of common AI models on all investigated CWEs, though the performance of FL-based vulnerability detection is limited by heterogeneous data.
%
%
To highlight the performance differences between different FL solutions for vulnerability detection, we extensively investigate the impacts of different configuration strategies for each framework component of VulFL, involving various mainstream data processing strategies, parameter-efficient model training schemes, and FL algorithms built on top of different Natural Language Processing (NLP)-based or Graph Neural Network (GNN)-based models. 
Our study sheds light on the potential of FL in vulnerability detection, which can be used to guide the design of FL-based solutions for vulnerability detection.

\end{abstract}



\maketitle

\section{Introduction}

As a promising software Quality Assurance (QA) technology, vulnerability detection \cite{wang2023ase,zhang2023ase,sejfia2023icse}
plays a vital role in ensuring the robustness and security of software projects.
However, due to the exponentially increasing complexity and scale of software, developers will have to invest significantly more time and effort in vulnerability detection, inevitably prolonging the agile development processes of modern software systems.
To mitigate this problem, more and more developers resort to Deep Learning (DL) techniques~\cite{sysevr,vuldeepecker,vulchecker,vuldeelocator}, especially Large Language Models (LLMs), to perform vulnerability detection, since they provide powerful code understanding and reasoning capabilities.
%
So far, existing  DL-based vulnerability detection methods can be mainly classified into two categories: 
i) Graph Neural Network (GNN)-based methods~\cite{devign,reveal,cao2023icse,wen2023icse,zhang2022ase} that can deal with various graph-based inputs including Abstract Syntax Trees (ASTs), Control Flow Graphs (CFGs) and Dataflow Graphs (DFGs);  
and ii) Text sequence neural network-based methods~\cite{vuldeepecker} for Natural Language Processing (NLP) tasks that take token sequences generated from text data as input.


Although DL-based vulnerability detection methods enable knowledge sharing among participating companies or organizations, they are facing the serious challenge of insufficient training data. 
Typically,  training a well-performing DL model requires a large amount of high-quality and diverse vulnerability data. 
However, few existing companies or organizations can maintain such a set of sufficient vulnerability data for model training.
Worse still, due to concerns about privacy leakage, most companies and organizations are reluctant to share their vulnerability data for collaborative model training.
All these factors result in the notorious {\it data silo problem},  
strongly limiting the use of DL techniques in vulnerability detection tasks. 
To address this issue, 
Federated Learning (FL)~\cite{mcmahan2017communication, hu2024kdd, FedProx, huang2023federated,hu2023gitfl} has been proposed as an effective distributed machine learning-based solution.
Based on a client-server architecture, FL maintains a global model on the cloud server for knowledge aggregation and sharing, enabling collaborative training among clients without compromising their data privacy. 
%
In each FL training round, the cloud server first dispatches the global model to clients for local model training and then aggregates the trained local models to update the global model.
Since only model gradients are transmitted between the cloud server and participant clients, 
the privacy of clients can be guaranteed. 

%

Due to the merits of privacy-aware collaborative learning, FL has been considered a new means for DL-based vulnerability detection.
As an initial attempt at FL-based vulnerability detection, the work VDBFL~\cite{vdbfl}
demonstrates the feasibility and potential of FL in enhancing the performance of vulnerability detection. 
However, it focuses on GNN-based performance optimization for vulnerability detection. 
Without taking various important issues such as i) DL models, ii) data heterogeneity scenarios, and iii) categories of vulnerabilities into account, existing FL-based vulnerability detection methods, including VDBFL, fail to comprehensively explore the potential and limitations of FL in dealing with common vulnerability detection tasks.
For example, although LLMs play an important role in NLP-based vulnerability detection, they have not yet been explored as underlying DL models in FL-based vulnerability detection. 
%
Besides the model issue, existing FL methods suffer from the ``gradient divergence'' problem ~\cite{icml_scaffold} caused by underlying data heterogeneity, which strongly degrades the overall collaborative learning performance of FL. 
Without a thorough study on this topic, developers will often get frustrated when contributing their heterogeneous code data for FL-based vulnerability detection.
Meanwhile, different code owners may encounter different types of Common Vulnerabilities and Exposures (CWEs), thus requiring that FL-based vulnerability detection methods can report a wide spectrum of CWEs. Unfortunately, existing FL-based vulnerability detection methods, including VDBFL, do not support this feature.  
So far, there is a lack of study to figure out {\it how well FL 
adapts to common vulnerability detection tasks}. 
%
%
Moreover, since different vulnerability detection methods adopt different data preprocessing and model training methods and numerous FL methods exist with various aggregation and communication strategies, it is hard for code owners to figure out one effective and efficient FL-based configuration for their specific vulnerability detection requirements. 
Clearly, to promote FL-based vulnerability detection, there is an urgent need to 
figure out {\it how to design a high-performance FL solution for a specific vulnerability detection task}.

To explore the above two issues, this paper presents a comprehensive study using our designed evaluation framework named {\it VulFed} for FL-based vulnerability detection, which consists of four key configurable components, i.e., the pre-processor, trainer, aggregator, and client selector.
By using VulFed, we first evaluate the performance of the vanilla FL method FedAvg~\cite{mcmahan2017communication} on a well-known real-world dataset  DiverseVul~\cite{diversevul}, based on a classic NLP-based model  CodeBERT~\cite{feng2020codebert}. 
To understand the effectiveness of FL in vulnerability detection, we compare FedAvg with independent training on four metrics and analyze the performance improvement on each CWE.
Then, we evaluate the performance of FedAvg within various non-IID scenarios to show the impact of data heterogeneity on the performance degradation of CWEs. 
In our study, we investigate the vulnerability detection performance of VulFed on top of 
three datasets, five data pre-processing techniques, six FL methods, nine DL models (six LLMs and three GNNs), and six 
model training schemes.
Comprehensive experimental results show that FL can achieve better inference accuracy on all CWEs than independent training, though different CWEs exhibit varying levels of robustness to data heterogeneity. Meanwhile, we can find that data heterogeneity is a non-negligible factor that strongly affects the overall inference accuracy. 
Since different component configurations strongly influence the overall performance of FL-based vulnerability detection, our results can help designers figure out an optimal FL solution for their specific requirements. 

In summary, the main contributions of this paper are as follows:
\begin{itemize}
\item
We propose a unified evaluation framework named VulFed for FL-based vulnerability detection, which can seamlessly integrate different  FL methods,  data pre-processing techniques, and model training schemes. 



\item 
We conduct extensive performance comparisons between VulFed and independent training to explore the potential and limitations of FL for common vulnerability detection tasks within both  Independent and Identically Distributed (IID) and non-IID scenarios.


\item 
We evaluate the performance of FL-based vulnerability detection with various VulFed component configurations, aiming to explore a strategy that can achieve a high-performance FL-based solution for a specific vulnerability detection task. 


\end{itemize}

The rest of this paper is organized as follows.
Section~\ref{sec:background} introduces the background of FL-based vulnerability detection.
Section~\ref{sec:overview} introduces the global design and detailed implementation of our evaluation framework.
Section~\ref{sec:studydesign} expounds on the study design coupled with three research questions and details the experimental evaluation settings. 
Section~\ref{sec:evaluation} presents all evaluation results to answer the three research questions.
After Section~\ref{sec:discussion} discusses the limitations of VulFed, Section~\ref{sec:conclusion} concludes the paper and reiterates our findings.
Finally, Section~\ref{sec:datable} provides channels for obtaining the data and source code of our framework.

\section{Background}
\label{sec:background}

\subsection{DL-based Vulnerability Detection}\label{sec:back_vul}
With the development of AI technologies, vulnerability detection methods based on DL exhibit powerful capabilities.
Existing DL-based vulnerability detection methods can be classified into two categories, i.e., NLP-based and GNN-based vulnerability detection.
NLP-based methods~\cite{vuldeepecker,mvuldeepecker,vuldeelocator} 
treat source codes as a special type of natural language.
They typically preprocess the source code into tokens and then utilize them to train NLP classification models.
For example, VulDeePecker~\cite{vuldeepecker} processes vulnerable C/C++ codes as code gadgets, i.e., text sequences, for training.
muVulDeePecker~\cite{mvuldeepecker} and VulDeeLocator~\cite{vuldeelocator} improve VulDeePecker by incorporating code attention, intermediate representation, and refined classification granularity into their methods.
VUDENC~\cite{vuldenc} modify VulDeePecker into a Python codes-oriented vulnerability detection model.
With the advantages of LLMs in logical reasoning, recent works~\cite{vuldebert, vulgpt, transformervul} adopt LLMs to replace the small NLP model for vulnerability detection.
Furthermore, SySeVR~\cite{sysevr} utilizes syntax and semantic information in control flow logic to construct more efficient candidate code slices for model training.
GNN-based methods adopt structural information within source codes, such as AST and CFG, to extract vulnerability features and then use GNN models for vulnerability detection.
For example, Devign ~\cite{devign} constructs a joint syntax tree based on AST and uses GNN to extract features before performing binary classification using convolutional layers.
Reveal~\cite{reveal} improves Devign by adopting GGNN to obtain graph embedding for CPG and optimizing the learning process. 

\subsection{Federated Learning}\label{sec:back_fl} 
Typically, an FL system consists of a cloud server and multiple clients.
In each FL training round, the cloud server dispatches a global model to local clients, and each client uses their raw data to train the received model.
Then, each client uploads its trained model to the cloud server, and the cloud server aggregates all the collected models to update the global model.
The objective of FL is commonly to optimize a global model represented as follows:
{
\setlength\abovedisplayskip{0.1cm}
\setlength\belowdisplayskip{0.1cm}
\begin{eqnarray}
\label{eq1}
\underset{\mathcal{W}}{min}F(\mathcal{W}) = \frac{1}{N}\sum_{i=1}^N\mathcal{L}_i(\mathcal{W};\mathcal{D}_i),
\end{eqnarray}
}
where $N$ denotes the number of clients, $\mathcal{W}$ is the global model's parameters,  $\mathcal{L}_i$ indicates the loss generated by each client, and $\mathcal{D}_i$ represents the local data provided by each client. 

The main challenge in FL is the data heterogeneity problem.
To address these problems, existing FL methods can be classified into five categories: global variable-based, clustering-based, knowledge distillation-based, mutation-based, and multi-model-based methods.
The global variable-based methods~\cite{icml_scaffold,fedcross} aim to use a global variable to guide the local models in clients to be optimized towards a similar direction.
Clustering-based methods~\cite{clusamp,li2021federated} cluster clients according to their specific characteristics, such as data distribution and activation of specific layers.
Based on clusters, the cluster-based methods prefer to fairly select clients from each cluster to participate in each round of local training.
Knowledge distillation-based methods~\cite{zhu2021data,zhang2022fine} conduct knowledge distillation technologies using a public or proxy dataset to optimize model training.
Multi-model-based methods~\cite{fedcross,hu2023fedmr} use multiple homogeneous models for local training rather than the same global model and use heuristic model collaboration methods to transform knowledge among models.
In this way, multi-model-based methods guide the models optimized towards a flat area, thereby achieving a well-generalized performance.
Mutation-based method~\cite{fedmut} mutates the global model to make the model shift in the solution space to eliminate the optimal local solution.

Recently, FL has been widely used in various applications, such as AIoT systems~\cite{chen2024flexfl,jia2024adaptivefl}, autonomous driving~\cite{zhang2021end}, and medical health~\cite{xu2021federated}.
However, FL-based vulnerability detection is still in the early stages.
VDBFL~\cite{vdbfl} presents an attempt for FL-based vulnerability detection.
However, VDBFL only explores the performance of GNN-based vulnerability detection using FL. At the same time, no more extensive research has been conducted, so it is still unclear how well the FL-based vulnerability detection using various models, especially LLMs, on different types of CWEs.
In addition, how to use FL wisely for vulnerability detection still needs further exploration.

\subsection{Training Scheme for Large Language Model}\label{sec:back_llm}
LLMs have demonstrated strong capabilities in both language understanding and logical reasoning.
However, since mainstream LLMs have hundreds of billions of parameters, the cost of LLM training is typically unacceptable for most companies or organizations.
To address this problem, the mainstream LLM-based methods typically adopt i) a public pre-trained model as the initial model and ii) a Parameter-Efficient Fine-Tuning (PEFT) strategy for model training.

Existing well-used LLMs can be categorized into three types based on their structure, i.e., encoder-only, decoder-only, and encoder-decoder structure.
For example, BERT~\cite{bert} is an encoder-only LLM oriented to text pair tasks, pre-trained through the two-step process of Mask Language Model (MLM) and Next Sentence Prediction (NSP).
CodeBERT~\cite{feng2020codebert} is a BERT-based LLM, which can handle code tasks better by adjusting pre-training data and structure design.
GPT~\cite{gpt} is a decoder-only LLM. The GPT series includes multiple iterative versions, and each newer version has better performance and a larger scale.
OPT~\cite{opt} is a series of LLMs with a structure similar to GPT but pre-trained by different data, with superior openness.
T5~\cite{t5} and CodeT5~\cite{codet5} utilize an encoder-decoder structure, whereas CodeT5 is an improved, code tasks-oriented T5-based model.

The vanilla Parameter-Efficient Fine-Tuning (PEFT)~\cite{PEFT} method froze most parameters in the LLM, referred to as the backbone, and only remains a small proportion of parameters for training, such as an external set of shallow networks or modules simplifying internal model computations.
Various PEFT schemes have recently been proposed, and the most well-used strategies are soft prompt and low-rank fine-tuning schemes.
For example, P-Tuning V1~\cite{P-tuning} and V2~\cite{P-tuningv2} are two mature paradigms within soft prompt tuning.
During training, trainable prompt vectors constructed before the input layer of the transformer block, along with additional MLP layers, serve as the hot parameters. In contrast, all internal parameters of the LLM constitute the backbone.
Low-rank Adaption (LoRA)~\cite{lora} is one of the most famous low-rank fine-tuning schemes, which involves injecting rank decomposition matrices as trainable parts into each block while freezing the weights within the model, thus effectively reducing the size of parameters to be learned.
LoHa~\cite{loha} and IA3~\cite{ia3} are both low-rank training schemes with different structural designs from LoRA. They perform better when facing specific scenarios.

\begin{table*}[t]
\caption{Details of PEFT Strategies Adopted in VulFed.}
\vspace{-0.1in}
\label{peft}
\centering
\resizebox{\textwidth}{!}{%
\begin{tabular}{@{}c|c|c|l@{}}
\hline\hline
PEFT Strategy & \multicolumn{1}{c|}{Type} & Hot Param. Rate & Description \\ \hline
No PEFT & Full-parameter & 100\% & if there is no PEFT strategy, all parameters will be trained. \\ \hline
P-Tuning V1~\cite{P-tuning} & Soft Prompt & 1.41\% & \begin{tabular}[c]{@{}l@{}}a method that adds a soft prompt to the original tokens to form new inputs and includes \\ an additional MLP layer.\end{tabular} \\ \hline
P-Tuning V2~\cite{P-tuningv2} & Soft Prompt & 10.60\% & an improved version of the V1, is also an enhanced iteration of prompt tuning. \\ \hline
LoRA~\cite{lora} & Low Rank & 0.23\% & \begin{tabular}[c]{@{}l@{}}Short for Low-Rank Adaption, which involves injecting rank decomposition matrices as \\ trainable parts into each block while freezing the weights within the model.\end{tabular} \\ \hline
LoHa~\cite{loha} & Low Rank & 1.29\% & \begin{tabular}[c]{@{}l@{}}a method that combines more low-rank matrices through the Hadamard product to approximate \\ the original large-scale matrix, similar to LoRA.\end{tabular} \\ \hline
IA3~\cite{ia3} & Low Rank & 0.05\% & \begin{tabular}[c]{@{}l@{}}an improved approach that follows the idea from LoRA of low-parameter fine-tuning within \\ the transformed internal structure of transformer blocks.\end{tabular} \\ \hline\hline
\end{tabular}
}
\vspace{-0.05in}
\end{table*}

\begin{table*}[t]
\caption{Details of FL Algorithms Adopted in VulFed.}
\vspace{-0.1in}
\label{fl}
\centering
\resizebox{\textwidth}{!}{%
\begin{tabular}{@{}c|c|c|l@{}}
\hline\hline
Algorithm & Type & Debut Year & Description \\ \hline
FedAvg~\cite{mcmahan2017communication} & Classic & 2017 & \begin{tabular}[c]{@{}l@{}}the most typical FL algorithm. It adopts a process of distributing models, local training, uploading gradients, \\ and global aggregation.\end{tabular} \\ \hline
FedProx~\cite{FedProx} & Global Variable & 2020 & \begin{tabular}[c]{@{}l@{}}an improved version of FedAvg which enhances performance by adding a proximal term to the original loss \\ function and increasing the tolerance of the cloud server towards locally insufficient training gradients.\end{tabular} \\ \hline
CluSamp~\cite{clusamp} &  Clustering & 2021 & \begin{tabular}[c]{@{}l@{}}an algorithm based on clustering, where clients are clustered and sampled based on sample size or model \\ similarity before training.\end{tabular} \\ \hline
FedCross~\cite{fedcross} &  Multi-Model & 2024 & \begin{tabular}[c]{@{}l@{}}an algorithm specifically designed to address non-IID data in FL by cross-aggregating effectively and \\ distributing intermediate model based on a distance judgment and using the global model only  for inference.\end{tabular} \\ \hline
\multirow{3}{*}{\shortstack{Moon~\cite{moon}}} & \multirow{3}{*}{\shortstack{Contrastive \\ Learning-based}} & \multirow{3}{*}{2021} & \multirow{3}{*}{\begin{tabular}[c]{@{}l@{}}an efficient FL algorithm based on contrastive learning. By introducing multiple representations to add  a   \\  model-contrastive loss during local training, Moon can control the training deviation of local models relative \\ to the global distribution.\end{tabular}} \\
&&& \\
&&& \\ \hline
FedMut~\cite{fedmut} & {Mutation} & 2024 & \begin{tabular}[c]{@{}l@{}}a method that mutates the global model based on gradient changes to generate multiple intermediate models \\ for the next round of training. Each intermediate model will be dispatched to the client for local training. \\ Finally, the global model converges to a flat region within the range of the  mutation model.\end{tabular} \\ \hline\hline
\end{tabular}
}
\vspace{-0.05in}
\end{table*}

\section{Our Evaluation Framework}
\label{sec:overview}

\begin{figure*}[h]
\centering
\includegraphics[width=0.99\textwidth]{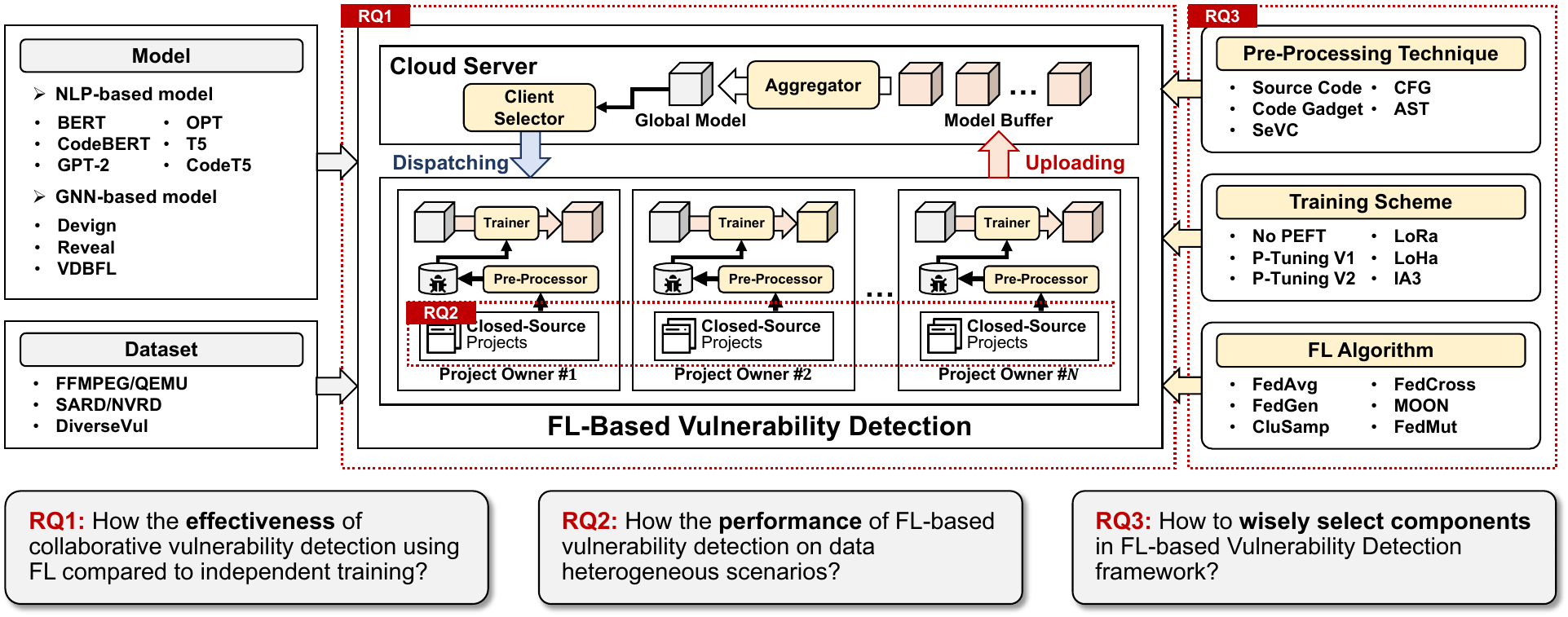}
 \vspace{-0.1in}
\caption{Our VulFed Evaluation Framework.} 
\label{frameraw}
 \vspace{-0.1in}
\end{figure*}

\subsection{Framework Design}
To understand the effectiveness of FL for vulnerability detection, according to the study in Section~\ref{sec:background},  we designed a general and easy-to-extend evaluation framework, named VulFed, which integrates various vulnerability detection methods and FL optimization methods.
As shown in Figure~\ref{frameraw}, VulFed adopts a classic central cloud-based FL architecture, consisting of a cloud server and multiple project owners.
To integrate various vulnerability detection and FL optimization methods, VulFed includes four configurable components, i.e., pre-processor, trainer, aggregator, and client selector, respectively.
Specifically, the pre-processor integrates data transformation methods to convert source code into a format suitable for specific DL-based vulnerability detection methods.
The trainer incorporates training schemes for specific models and optimization strategies for FL methods.
The aggregator implements model aggregation and server-side optimization strategies for FL methods.
The client selector employs client selection strategies tailored to specific FL methods.
By configuring the four components, we can implement vulnerability detection methods based on different FL optimization algorithms.

\subsection{Framework Implementation}
\subsubsection{Pre-processing techniques}
As described in Section~\ref{sec:back_vul}, existing vulnerability detection methods are based on different data structures.
To integrate various vulnerability detection methods on VulFed, we implemented our pre-processer to convert source code into four data structures, i.e., code gadget, SeVC, CFG, and AST, where the first two structures are used for LLM-based vulnerability detection methods and the last two structures are used for GNN-based vulnerability detection methods.
Note that VulFed supports directly using source code for model training.

\subsubsection{Training scheme}
Due to the limited hardware resources available in clients, the communication and computing overhead of the full-parameter training of LLM is unacceptable in FL.
As shown in Table~\ref{peft}, to effectively evaluate the performance of LLM-based vulnerability detection, the trainer of FedVul integrates six training schemes,i.e, full-parameter, P-Tuning V1~\cite{P-tuning}, P-Tuning V2~\cite{P-tuningv2}, LoRA~\cite{lora}, LoHa~\cite{loha}, and IA3~\cite{ia3}, where P-Tuning V1 and V2 are soft prompt-based PEFT schemes, while LoRA, LoHa, and IA3 are low rank-based PEFT schemes.

\subsubsection{FL algorithm}
To understand the performance of different FL algorithms in vulnerability detection, VulFed integrates six different types of FL algorithms, i.e., FedAvg~\cite{mcmahan2017communication}, FedProx~\cite{FedProx}, CluSamp~\cite{clusamp}, FedCross~\cite{fedcross}, MOON~\cite{moon}, and FedMut~\cite{fedmut}, respectively.
Table~\ref{fl} presents the type and description of each algorithm.
Note that, since different optimization algorithms involve different processes in FL, VulFed needs to configure different components for each algorithm.
For example, since FedProx needs to use a proximal tern to regular local training, VulFed needs to configure the trainer.
Since CluSamp needs to cluster clients into multiple groups and select clients according to groups, VulFed should configure the aggregator and client selector.
Note that VulFed can easily integrate FL algorithms by implementing different configuration components.

\section{Study Design}
\label{sec:studydesign}

\subsection{Overview and Research Questions}
We conducted extensive experiments to evaluate the performance of VulFed involving various vulnerability detection tasks. 
Our objectives are to i) understand the effectiveness of different FL components in detecting vulnerabilities, focusing on the  performance of detecting vulnerabilities of different categories by different DL models within various data heterogeneity scenarios,
and ii) help code owners manage the underlying FL framework to maximize the performance of collaborative vulnerability detection, e.g., provide useful guidance on selecting the best suitable FL components or settings according to the specific requirements and contexts of given vulnerability detection tasks. All these experiments try to answer the following three
research questions (RQs). 
%


{\bf RQ1: How effective is collaborative vulnerability detection using FL compared to independent training?} We conducted performance comparisons on various vulnerability detection tasks, aiming to figure out the superiority of FL over traditional independent training in vulnerability detection. In other words, we want to figure out can the collaborative training scheme provided by FL can substantially improve the vulnerability detection performance achieved by existing methods that rely on independent learning on local data only. 
%
To evaluate the effectiveness of FL in detecting different types of vulnerabilities, we analyzed performance improvements within the subsets of samples corresponding to different CWE categories.
%


{\bf RQ2: How effective is FL-based vulnerability detection on data heterogeneity scenarios?}
Due to differences in business and specific technology stacks, the distributions of vulnerability data of different project owners vary significantly.
To investigate the effectiveness of FL-based vulnerability detection on data heterogeneity scenarios, we simulated non-IID situations by constructing heterogeneous client data using the Dirichlet distributions~\cite{dirichlet}. We evaluated the model performance within various simulated heterogeneous scenarios and conducted statistical analyses for both binary tasks and multi-classification tasks according to their CWE types.


{\bf RQ3: How to wisely select components in FL-based vulnerability detection frameworks?}
To figure out optimal configuration strategies for different components of an FL-based vulnerability detection framework, we constructed VulFed by integrating various forms of code inputs into a flexible FL framework with multiple customizable components. 
Specifically, VulFed supports adjusting the deployed models, fine-tuning strategies, and FL algorithms by controlling the configuration options. 
Based on  VulFed, we conducted extensive experiments on three types of code samples, exploring the impacts of different component combinations of the FL-based vulnerability detection framework within both IID and non-IID situations.




\begin{table}[h]
\caption{Statistics of Datasets.}
\vspace{-0.1in}
\label{dataset}
\centering
\footnotesize
\begin{tabular}{c|c|c|c|c}
\hline\hline
Dataset Name & Data Form & Code Source & Data Size & Vulnerability Size \\ \hline
Devign & Source Code & FFMPEG/QEMU & 24586 & 11273 \\ \hline
VulDeePecker & Code Gadget & \multirow{2}{*}{SARD/NVRD} & 39753 & 10440 \\ \cline{1-2} \cline{4-5} 
SySeVR & SeVCs &  & 22062 & 5639 \\ \hline
DiverseVul & Multi-Classes Source Code & Real-World Projects & 328436 & 16889 \\ \hline\hline
\end{tabular}
\vspace{-0.1in}
\end{table}

\subsection{Experimental Settings}

{\bf Dataset Settings.}
We investigated four well-known datasets from three different data sources to conduct our experiments. Table~\ref{dataset} presents the detailed information of the datasets.
Specifically, the data utilized for training all GNN-based methods and the LLM without preprocessing steps is obtained from FFMPEG, a multimedia library, and QEMU, a hardware virtualization emulator. The data in the form of Code Gadget and SeVC were generated from SARD/NVRD, and we directly employed the datasets constructed in~\cite{vuldeepecker} and~\cite{sysevr} during the training and evaluation phases. Furthermore, for the fine-grained classification task related to CWE types, we conducted training on a more realistic and meticulously annotated dataset sourced from ~\cite{diversevul}. Note that 
we directly adopted the preset division of a dataset if it has been divided. Otherwise, we divided the dataset by 80\% as the training set and 20\% as the testing set.

{\bf Model Settings.}
We considered six different LLMS, i.e.,  BERT, CodeBERT, GPT-2, OPT, T5, and codet5 as deployable models. We used the lightweight versions of all the above models to better adapt to FL scenarios with a small computational overhead. Table~\ref{model} details all the adopted LLMs.

\begin{table*}[t]
\caption{Details of Deployable LLMs used in VulFed.}
\vspace{-0.1in}
\label{model}
\centering
\resizebox{\textwidth}{!}{%
\begin{tabular}{@{}c|c|c|c|l@{}}
\hline\hline
Model & \multicolumn{1}{c|}{Scale} & \multicolumn{1}{c|}{Memory Usage} & Parameter \# & Description \\ \hline
BERT~\cite{bert} & Base & 440.5M & 110M & \begin{tabular}[c]{@{}l@{}}an encoder-based llm with a dual-sentence input structure, accomplished through\\ joint pre-training tasks of Masked Language Model and Next Sentence Prediction.\end{tabular} \\ \hline
CodeBERT~\cite{feng2020codebert} & Base & 498.6M & 110M & \begin{tabular}[c]{@{}l@{}}built upon RoBERTa, incorporates a substantial amount of code datasets for\\  additional pre-training.\end{tabular} \\ \hline
GPT-2~\cite{gpt2} & Small & 548.1M & 117M & \begin{tabular}[c]{@{}l@{}}a decoder-based LLM pre-trained unsupervised on raw text. It possesses several\\ aspects of improvements compared to GPT.\end{tabular} \\ \hline
OPT~\cite{opt} & Small & 250.5M & 125M & \begin{tabular}[c]{@{}l@{}}a class of open-source high-quality large models. It adopts a similar decoder\\ architecture to GPT.\end{tabular} \\ \hline
T5~\cite{t5} & Small & 242.1M & 60M & \begin{tabular}[c]{@{}l@{}}short for Text-to-Text Transfer Transformer, which is based on the principle of \\ transforming all NLP tasks into a text-to-text format that can be solved using\\ a unified model.\end{tabular} \\ \hline
CodeT5~\cite{codet5} & Small & 242.0M & 60M & an optimized version of the T5 model tailored for code-related tasks. \\ \hline\hline
\end{tabular}
}
\vspace{-0.1in}
\end{table*}

{\bf Metrics.}
To accurately measure the classification performance of the model, we used the four most commonly used evaluation criteria for vulnerability detection tasks as our metrics, i.e.,  accuracy, precision, recall, and F1 score. Precision and recall reflect the tendency of the model to represent positive or negative samples to some extent, while accuracy and F1 score, as more balanced and comprehensive metrics, have greater reference values.

{\bf Implementation Details.}
We implemented our VulFed framework using PyTorch. All the experimental results were obtained from an Ubuntu workstation equipped with an NVIDIA GeForce RTX 4090 GPU and an Intel i7-13700K CPU. All items of our evaluation are the test results of the classification performance of the model after 50 rounds of joint fine-tuning in an FL system with 10 clients. All clients and servers adopted models of the same type. In each round of training, 50\% of the clients were selected to participate in FL training with a local learning rate of $1\times e^{-5}$.

\section{Evaluation Results}
\label{sec:evaluation}

\subsection{RQ1: The Effectiveness of Using FL}
\label{effective}

\begin{table*}[t]
\caption{Details of DiverseVul after Pre-processing}
\label{cwecase}
\vspace{-0.1in}
\centering
\scriptsize
\begin{tabular}{c|c|l}
\hline\hline
CWE Type & Sample $\#$ & Description \\ \hline
Security Sample & 330492 & samples with no vulnerability \\ \hline
CWE-20 & 1315 & improper input validation \\ \hline
CWE-22 & 140 & improper limitation of a pathname to a restricted directory (`Path Traversal') \\ \hline
CWE-59 & 165 & improper link resolution before file access (`Link Following') \\ \hline
CWE-119 & 1435 & improper restriction of operations within the bounds of a memory buffer \\ \hline
CWE-120 & 285 & buffer copy without checking size of input (`Classic Buffer Overflow') \\ \hline
CWE-125 & 1635 & out-of-bounds read \\ \hline
CWE-189 & 275 & numeric errors \\ \hline
CWE-190 & 675 & integer overflow or wraparound \\ \hline
CWE-200 & 715 & exposure of sensitive information to an unauthorized actor \\ \hline
CWE-264 & 215 & permissions, privileges, and access controls \\ \hline
CWE-269 & 110 & improper privilege management \\ \hline
CWE-284 & 285 & improper access control \\ \hline
CWE-287 & 105 & improper authentication \\ \hline
CWE-295 & 100 & improper certificate validation \\ \hline
CWE-310 & 345 & cryptographic issues \\ \hline
CWE-362 & 400 & concurrent execution using shared resources with improper synchronization (`Race Condition') \\ \hline
CWE-369 & 160 & divide by zero \\ \hline
CWE-399 & 460 & resource management errors \\ \hline
CWE-400 & 395 & uncontrolled resource consumption \\ \hline
CWE-401 & 195 & missing release of memory after effective lifetime \\ \hline
CWE-415 & 245 & double free \\ \hline
CWE-416 & 1000 & use after free \\ \hline
CWE-476 & 915 & NULL pointer dereference \\ \hline
CWE-617 & 160 & reachable assertion \\ \hline
CWE-703 & 735 & improper check or handling of exceptional conditions \\ \hline
CWE-772 & 100 & missing release of resource after effective lifetime \\ \hline
CWE-787 & 1380 & out-of-bounds write \\ \hline
CWE-835 & 110 & loop with unreachable exit condition (`Infinite Loop') \\ \hline
CWE-None & 2835 & samples from commit associated without any CWE type \\ \hline\hline
\end{tabular}
\end{table*}

To evaluate the effectiveness of FL in vulnerability detection, we compared the performance of vanilla FL-based with independent training-based vulnerability detection methods in VulFed, where each client, in the latter case, only utilizes its raw data to train an isolated local model.
Here, we selected the DiverseVul dataset for evaluation because it is a relatively new and comprehensive dataset among those deployed in VulFed, encompassing a sufficient amount of vulnerable code and its corresponding CWEs.
Given that DiverseVul originates from various authentic sources of real-world vulnerability reports, the distribution of its labels can be considered representative of real-world scenarios.
We cleaned the dataset by retaining all secure code samples and removing CWE categories with fewer than 100 data samples.
It should be noted that the ratio of positive to negative samples in the preprocessed dataset nearly remained unchanged after implementing the schemes above since the volume of the discarded samples was rare.
Table~\ref{cwecase} shows the detailed information of the preprocessed DiverseVul dataset.
This dataset includes 30 categories comprising general CWE types, the non-vulnerable sample type, and the type of unspecified CWE label.
Compared to traditional vulnerability detection methods, which are generally based on a binary classification task, we chose to evaluate the inference performance of our trained models for each CWE type to explore the effectiveness of FL on different types of vulnerability.

We intended to ascertain the potential association between the FL framework and samples belonging to each type of CWE and whether deploying the FL framework can help to find vulnerabilities better.
However, utilizing different deployed models will introduce varying biases in the evaluation results because the inference result of LLMs exhibits different tendencies based on the corpus content and strategies of the pre-training phase.
To eliminate bias introduced by pre-trained models, we implemented our evaluation on all kinds of lightweight LLMs deployed in our frameworks by comparing the detection performance between training the model independently and training the model under the FL framework when the client utilizes each same model.
The analysis of this RQ will be conducted on the average of these evaluation results.

We simulated 10 clients that participated in FL training when implementing our evaluation.
Meanwhile, we divided the entire dataset into ten parts with the same number of samples following the same distribution. 
Each client holds one of these data subsets as its local private data.
While conducting FL training on all clients to obtain a global evaluation model, we implemented different training modes as a control group to verify the effectiveness of FL training.
Specifically, to explore how much improvement FL training can bring compared to independent local training, we set up an extra isolated client.
This client possesses the same amount and distribution of private data as other clients but does not participate in FL training and only utilizes its private data to train an independent model for evaluation.

\begin{figure}[t]
	\centering
	\includegraphics[width=0.93\linewidth]{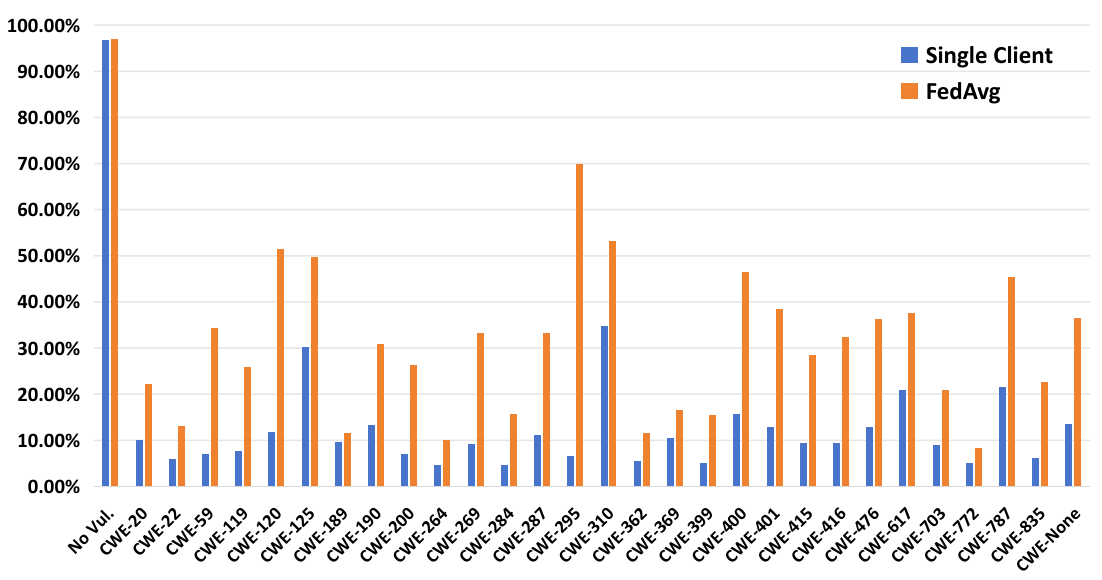}
\vspace{-0.10in}
    \caption{Performance Comparison between Single and FL Models}
\vspace{-0.10in}
	\label{rq1_fli}
 \vspace{-0.10in}
\end{figure}

Figure~\ref{rq1_fli} shows the differences in performance improvement of code samples belonging to different CWE types in FL, which are the average results evaluated by six different models deployed under our framework. 
Our direct finding from Figure~\ref{rq1_fli} is that the model trained under the FL framework has varying degrees of improvement in detecting vulnerabilities belonging to all CWE types compared to the model trained locally.
Meanwhile, the detection of non-vulnerable codes did not produce more false positives but remained at the original level, with a slight improvement.
Specifically, comparing the detection performance of the FL training model with the single client training model, the overall F1 Score increased from 18.33\% to 38.82\%.
On the other hand, we conducted collaborative training without deploying the FL framework to obtain the theoretically optimal performance of the global model.
Our FL model only produces less than 5\% loss on the F1 score compared to this public-trained global model.
The above findings indicate that, for vulnerability detection tasks, deploying FL frameworks to cover larger training scales to obtain a better global model is an effective strategy that can significantly improve the ability of the model to detect vulnerabilities.
However, a noticeable issue is that the improvement of detection performance of the model for code samples belonging to different CWE types varies significantly after incorporating FL training.
It is important to identify the underlying factors behind the differences in detection improvement rates between different vulnerability classes under VulFed.

By analyzing the detection rate of each type of vulnerability, we can discover which kinds of vulnerabilities are easier to detect and which ones are still difficult to find within the FL framework.
Table~\ref{rq1cwetable} shows the average performance comparison between independent training and FL training and ranks them according to the performance improvement level corresponding to each type of vulnerability.
It can be observed that there is a significant difference in the improvement of the detection ability of the model for vulnerabilities belonging to different CWE types after incorporating FL training.
The improvement value exceeded 25\% among CWE-401, CWE-295, etc., while on the other hand, the improvement generated among CWE-189, CWE-22, etc., was less than 10\%, or even almost none.
The general factor affecting the above results is generally considered to be the number of samples.
However, from Table~\ref{cwecase}, we can find that the correlation between the sample size of each type of vulnerability and the detection ability of the model for that type of vulnerability is insufficient.
For example, CWE-189 and CWE-295 are the types with the lowest and highest improvements in detection rates, respectively, but their sample sizes in DiverseVul are very similar, and the latter is even smaller than the former.
In addition, CWE-125 has the largest number of samples, but the improvement in detection performance for this type is still not as good as CWE-120 or CWE-400, which have much fewer samples.

\begin{table*}[t]
\caption{Evaluation results and improvement of different training strategies}
\label{rq1cwetable}
\vspace{-0.1in}
\centering
\resizebox{\textwidth}{!}{
\begin{tabular}{c|cccccccccccccc}
\hline\hline
\multirow{2}{*}{\begin{tabular}[c]{@{}c@{}}Performance\\ Type\end{tabular}} & \multicolumn{14}{c}{CWE ID} \\ \cline{2-15} 
 & \multicolumn{1}{c|}{189} & \multicolumn{1}{c|}{772} & \multicolumn{1}{c|}{264} & \multicolumn{1}{c|}{369} & \multicolumn{1}{c|}{362} & \multicolumn{1}{c|}{22} & \multicolumn{1}{c|}{399} & \multicolumn{1}{c|}{284} & \multicolumn{1}{c|}{703} & \multicolumn{1}{c|}{20} & \multicolumn{1}{c|}{835} & \multicolumn{1}{c|}{617} & \multicolumn{1}{c|}{190} & 119 \\ \hline
\begin{tabular}[c]{@{}c@{}}Independent\\ Training\end{tabular} & \multicolumn{1}{c|}{9.70\%} & \multicolumn{1}{c|}{5.00\%} & \multicolumn{1}{c|}{4.65\%} & \multicolumn{1}{c|}{10.42\%} & \multicolumn{1}{c|}{5.42\%} & \multicolumn{1}{c|}{5.95\%} & \multicolumn{1}{c|}{5.07\%} & \multicolumn{1}{c|}{4.68\%} & \multicolumn{1}{c|}{9.07\%} & \multicolumn{1}{c|}{10.14\%} & \multicolumn{1}{c|}{6.07\%} & \multicolumn{1}{c|}{20.83\%} & \multicolumn{1}{c|}{13.34\%} & 7.78\% \\ \hline
\begin{tabular}[c]{@{}c@{}}FL\\ Training\end{tabular} & \multicolumn{1}{c|}{11.52\%} & \multicolumn{1}{c|}{8.33\%} & \multicolumn{1}{c|}{10.08\%} & \multicolumn{1}{c|}{16.67\%} & \multicolumn{1}{c|}{11.67\%} & \multicolumn{1}{c|}{13.09\%} & \multicolumn{1}{c|}{15.58\%} & \multicolumn{1}{c|}{15.79\%} & \multicolumn{1}{c|}{20.86\%} & \multicolumn{1}{c|}{22.18\%} & \multicolumn{1}{c|}{22.73\%} & \multicolumn{1}{c|}{37.50\%} & \multicolumn{1}{c|}{30.86\%} & 25.79\% \\ \hline
\begin{tabular}[c]{@{}c@{}}Performance\\ Improvement\end{tabular} & \multicolumn{1}{c|}{1.82\%} & \multicolumn{1}{c|}{3.33\%} & \multicolumn{1}{c|}{5.43\%} & \multicolumn{1}{c|}{6.25\%} & \multicolumn{1}{c|}{6.25\%} & \multicolumn{1}{c|}{7.14\%} & \multicolumn{1}{c|}{10.51\%} & \multicolumn{1}{c|}{11.11\%} & \multicolumn{1}{c|}{11.80\%} & \multicolumn{1}{c|}{12.04\%} & \multicolumn{1}{c|}{16.67\%} & \multicolumn{1}{c|}{16.67\%} & \multicolumn{1}{c|}{17.53\%} & 18.01\% \\ \hline\hline
 & \multicolumn{1}{c|}{310} & \multicolumn{1}{c|}{415} & \multicolumn{1}{c|}{200} & \multicolumn{1}{c|}{125} & \multicolumn{1}{c|}{287} & \multicolumn{1}{c|}{416} & \multicolumn{1}{c|}{476} & \multicolumn{1}{c|}{787} & \multicolumn{1}{c|}{269} & \multicolumn{1}{c|}{401} & \multicolumn{1}{c|}{59} & \multicolumn{1}{c|}{400} & \multicolumn{1}{c|}{120} & 295 \\ \hline
\begin{tabular}[c]{@{}c@{}}Independent\\ Training\end{tabular} & \multicolumn{1}{c|}{34.78\%} & \multicolumn{1}{c|}{9.52\%} & \multicolumn{1}{c|}{6.99\%} & \multicolumn{1}{c|}{30.27\%} & \multicolumn{1}{c|}{11.11\%} & \multicolumn{1}{c|}{9.50\%} & \multicolumn{1}{c|}{12.93\%} & \multicolumn{1}{c|}{21.61\%} & \multicolumn{1}{c|}{9.09\%} & \multicolumn{1}{c|}{12.82\%} & \multicolumn{1}{c|}{7.07\%} & \multicolumn{1}{c|}{15.61\%} & \multicolumn{1}{c|}{11.69\%} & 6.67\% \\ \hline
\begin{tabular}[c]{@{}c@{}}FL\\ Training\end{tabular} & \multicolumn{1}{c|}{53.14\%} & \multicolumn{1}{c|}{28.57\%} & \multicolumn{1}{c|}{26.34\%} & \multicolumn{1}{c|}{49.64\%} & \multicolumn{1}{c|}{33.33\%} & \multicolumn{1}{c|}{32.50\%} & \multicolumn{1}{c|}{36.25\%} & \multicolumn{1}{c|}{45.41\%} & \multicolumn{1}{c|}{33.33\%} & \multicolumn{1}{c|}{38.46\%} & \multicolumn{1}{c|}{34.35\%} & \multicolumn{1}{c|}{46.41\%} & \multicolumn{1}{c|}{51.46\%} & 70.00\% \\ \hline
\begin{tabular}[c]{@{}c@{}}Performance\\ Improvement\end{tabular} & \multicolumn{1}{c|}{18.35\%} & \multicolumn{1}{c|}{19.05\%} & \multicolumn{1}{c|}{19.35\%} & \multicolumn{1}{c|}{19.37\%} & \multicolumn{1}{c|}{22.23\%} & \multicolumn{1}{c|}{23.00\%} & \multicolumn{1}{c|}{23.31\%} & \multicolumn{1}{c|}{23.79\%} & \multicolumn{1}{c|}{24.24\%} & \multicolumn{1}{c|}{25.64\%} & \multicolumn{1}{c|}{27.28\%} & \multicolumn{1}{c|}{30.80\%} & \multicolumn{1}{c|}{39.77\%} & 63.33\% \\ \hline\hline
\end{tabular}
}
\vspace{-0.05in}
\end{table*}

We then analyze the causation of the above phenomenon based on the actual situations of several CWEs.
It should be noted that the analysis we can perform only includes the qualitative part and cannot effectively construct a standard to quantify the performance improvement of target code categories in the FL framework because multiple factors simultaneously constitute the phenomenon to be analyzed.
Although we believe that the nature of the CWE type to which the code belongs is the biggest factor affecting the extent of improvement, the distribution tendency generated during the construction process of the dataset, and the randomness in model training even if it has been mitigated and avoided as much as possible, can also be one of the factors.
Therefore, we would rather focus on the CWE types that perform the best and worst and analyze them to draw a conclusion that applies in most cases.

\begin{figure}[htbp]
\vspace{-0.05in}
	\centering
	\includegraphics[width=\linewidth]{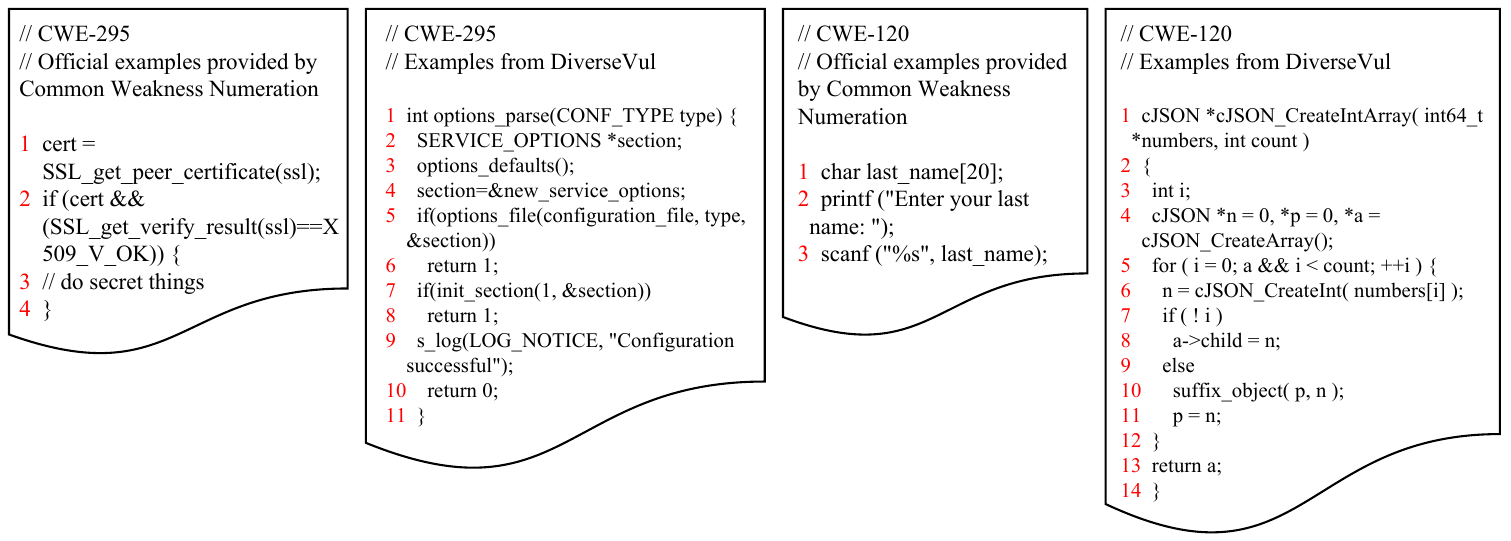}
 \vspace{-0.25in}
    \caption{Examples of CWE-295 and CWE-120 provided by the official and in  DiverseVul}
	\label{bettercode}
\end{figure}

\begin{figure}[htbp]
\vspace{-0.1in}
	\centering
	\includegraphics[width=\linewidth]{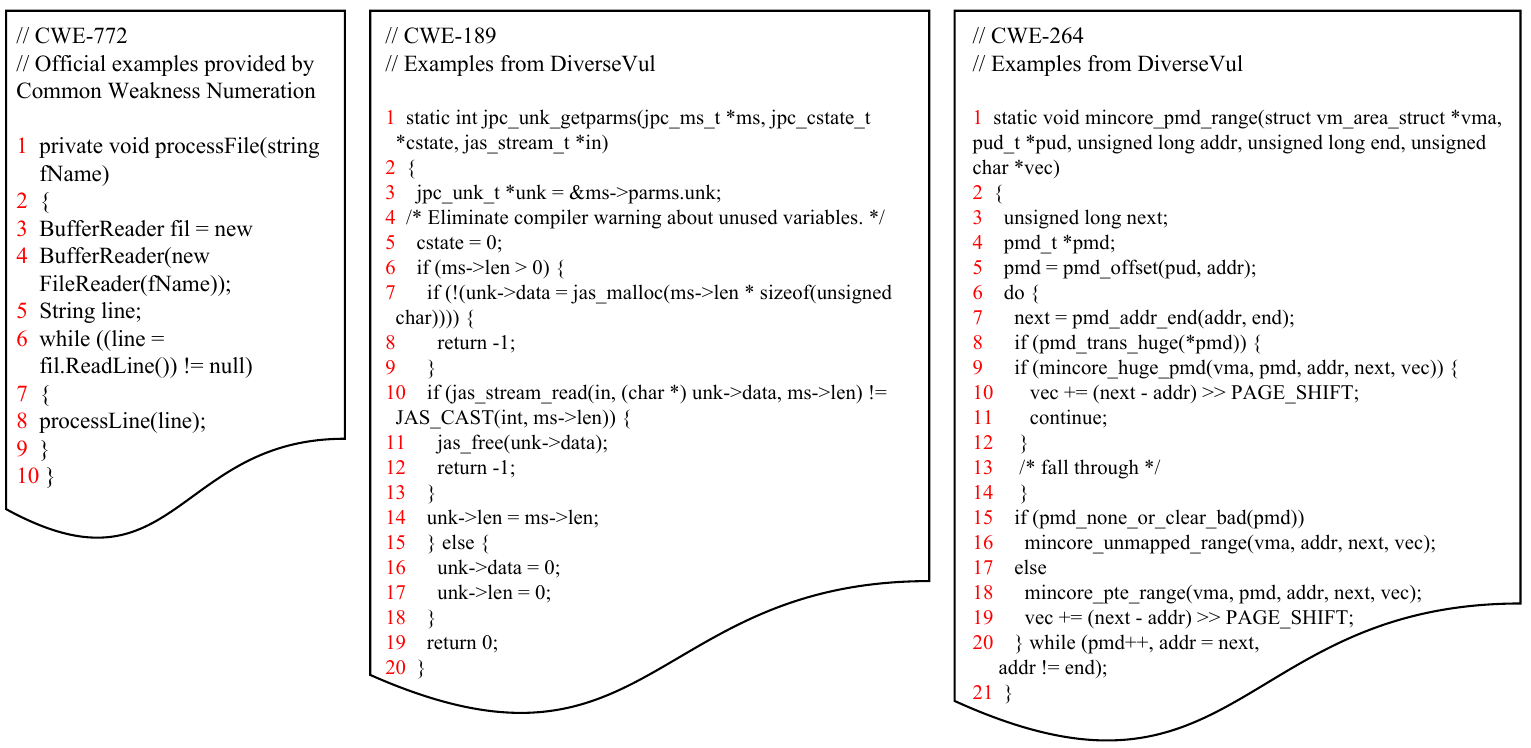}
    \vspace{-0.25in}
    \caption{Examples of CWE-772, CWE-189 and CWE-264 provided by the official and in the DiverseVul}
	\label{worsecode}
\end{figure}

Figure~\ref{bettercode} shows examples of two types of vulnerabilities, including CWE-295: ``Improper Certificate Validation'', and CWE-120: ``Buffer Copy without Checking Size of Input (`Classic Buffer Overflow')'', provided by official sources and appearing in the DiverseVul dataset.
These two CWE Types achieved performance improvements of 63.33\% and 39.77\%, respectively, under the FL framework.
Meanwhile, Figure~\ref{worsecode} shows examples of three types of vulnerability, including CWE-189: ``Numeric Errors'', CWE-772: ``Missing Release of Resource after Effective Lifetime'', and CWE-264: ``Permissions, Privileges, and Access Controls''.
The performance improvements corresponding to these three CWE types are all less than 6\%.

A direct conclusion that can be drawn from the above examples is that the shallower the vulnerability characterization properties of a certain vulnerability type itself, the lack of code areas involved, the easier the model can learn its knowledge, and the better the improvement of its classification performance under the FL framework.
On the contrary, vulnerabilities involving more different representations or possessing more vulnerable areas can result in less improvement with knowledge aggregation and an increase of data volume brought about by joining the FL framework.

Specifically, CWE-189 is a type of vulnerability with complex representation forms, and it is a starting sequence for many branch vulnerability types.
Meanwhile, CWE-264 is also a type of vulnerability representation associated with multiple branches.
The cases in CWE-772 may have higher complexity and involve more code areas, which increases the difficulty for the model in constructing features.
All the factors mentioned above are the reason for the poor classification performance and poor improvement in FL training.
Although vulnerabilities belonging to CWE-295 are business-oriented, the common features of vulnerabilities are highly correlated.
Once the data volume becomes sufficient, the model can quickly learn the common knowledge of this type of vulnerability.
The same applies to CWE-120 and other well-performed CWE types.
For vulnerability types that are difficult to improve in the model detection ability through FL training, other preprocessing methods should be utilized, such as designing data forms that enhance their semantic logic or transforming them into more unified intermediate representations.

\begin{tcolorbox}
    [
    colback=yellow!7!white,
    colframe=white!75!white,
    width=\textwidth,
    arc=1mm, auto outer arc,
    boxrule=0.5pt,
    ]
{\bfseries Answer to RQ1:}
The comprehensive performance of the VulFed model is much higher than that of models trained independently by the client, achieving twice the lead in the F1 Score, indicating that FL-based vulnerability detection is more effective.
However, there are differences in the detection ability of the model for samples belonging to different CWE types.
After joining FL training, types with simpler and more obvious vulnerability representations are more easily detected. In contrast, detecting types with more situations and higher complexity does not show effective improvement.

\end{tcolorbox}

\subsection{RQ2: Performance on Non-IID Scenarios}
\label{non-iid}

In the evaluation process of the previous research question, we utilized a relatively uniform data distribution when allocating the data for FL training.
However, it is well known that FL frameworks in practical applications inevitably suffer from heterogeneity among different clients, which primarily manifests in terms of data quality and data distribution.
On one hand, each client would commit a different number of vulnerabilities based on its own program business scale.
On the other hand, vulnerabilities committed by various clients may also follow different label distributions, which are related to many factors, including the business tendencies of enterprises or departments and the personal habits of code writers.
Actually, the vulnerability codes generated between clients may have inconsistent or even completely different vulnerability labels in the entire FL system.
Clients who possess more data can generally train more comprehensive local models.
In contrast, clients who only possess a small amount of data find it difficult for local models to acquire sufficient knowledge.
In addition, clients whose obtained data conforms to a more uneven label distribution will inevitably develop a bias towards classification during the training process.
The above issues may become key factors affecting the performance of the global model in the server aggregation part of FL training.
Therefore, it is necessary to evaluate what changes will appear in the vulnerability detection capability of the model trained under this framework when the non-IID scenarios mentioned above occur in VulFed.
In addition, similar to RQ1, we also hope to understand the resistance of the code samples corresponding to each CWE type to the data interference caused by non-IID scenarios.
Therefore, we conducted an experimental evaluation of VulFed under the premise of data heterogeneity and compared it with the case of uniform data distribution.

To simulate the problem of data heterogeneity during the experiments, we generally allocated different data sizes and label distributions from the global dataset to each client as its privacy data.
In the VulFed, we utilized a non-IID dataset partitioning method based on the Dirichlet distribution~\cite{dirichlet} to achieve heterogeneity.
During evaluating this RQ, the hyperparameter $\alpha$ of the Dirichlet distribution will be set to 0.5 by default.
When this parameter is set to a smaller value (e.g., 0.3 or 0.1), it does not significantly impact the experimental results based on our multiple verifications.
After simulating local data acquisition on clients, we conducted evaluations based on the multi-classification of code samples belonging to different CWE types to derive straightforward and representative conclusions to this RQ.
To analyze the data heterogeneity problem under the multi-classification vulnerability detection task, we continued to use the cleaned DiverseVul dataset described in~\ref{effective} as the global dataset.

\begin{figure}[t]
	\centering
	\includegraphics[width=0.8\linewidth]{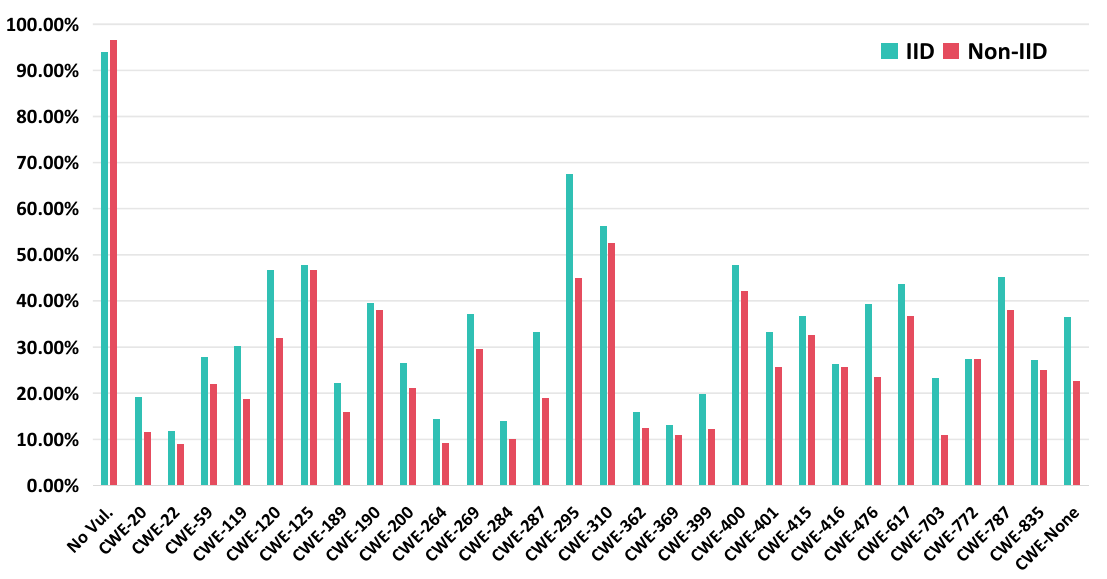}
\vspace{-0.10in}
    \caption{Performance Comparison between IID and non-IID Situations}
\vspace{-0.10in}
	\label{rq2_iid}
 \vspace{-0.1in}
\end{figure}

Multiple FL algorithms have been deployed in VulFed, and most have better resistance to data heterogeneity.
However, the differences in the nature of these methods will lead to variations in the detection ability of the model for each type of sample in the evaluation results.
Therefore, the scope of our evaluation in this section includes the models trained using each FL algorithm.
Figure~\ref{rq2_iid} shows the comparison between the overall average detection performance of all FL algorithms deployed in VulFed under non-IID situations and the detection performance under IID situations.
When evaluating the detection performance of the model utilizing each FL algorithm, multiple data allocations and training were performed in non-IID situations to obtain more realistic results.

The conclusion observed from Figure~\ref{rq2_iid} is that in non-IID situations, the false positive rate for detecting non-vulnerable samples remains equal and does not undergo significant changes.
On the other hand, the detection rates corresponding to all CWE types have decreased, which is an inevitable phenomenon caused by data heterogeneity.
Compared with the general situation of deploying non-IID data distributions in the field of FL, the accuracy reduction of various CWE types in VulFed is not remarkable; almost all of them remain within 15\%.
When confronting such complex data situations in vulnerability detection scenarios, such performance is acceptable.
Finally, from a macro perspective, the overall accuracy loss is less than 2\%, while the F1 Score loss is less than 5\%, which indicates that, considering the situations comprehensively when utilizing each kind of FL algorithm, VulFed has strong resistance to the problem of data heterogeneity.
Its performance in non-IID situations is similar to that in IID situations.

The phenomenon above is related to the strategy of employing LLMs.
The model has already learned from a large corpus during the pre-training phase, including corpora from the code domain, which enhances its performance floor in incremental learning.
Even when the training data is insufficient in quantity and quality, the directional deviation introduced by this local model during global aggregation remains within an acceptable range.
On the other hand, with a certain range of improvement in incremental data, the adaptation of the model to the domain will further increase.
Under the same total data volume, the strategy of the non-IID partitioning to the dataset also leads to higher-quality clients, which further benefits LLMs compared to traditional models.
Therefore, in almost all cases, the vulnerability detection ability of the model for samples belonging to each CWE type will not be significantly reduced, which maintains the metric at the same level.

\begin{tcolorbox}
    [
    colback=yellow!7!white,
    colframe=white!75!white,
    width=\textwidth,
    arc=1mm, auto outer arc,
    boxrule=0.5pt,
    ]
{\bfseries Answer to RQ2:}
VulFed is a robust framework against non-IID problems, exhibiting strong resistance to data heterogeneity issues. The performance of the model in non-IID scenarios is already close to that in IID scenarios. Samples belonging to all CWE types have performance degradation but are controlled within 15\%.

\end{tcolorbox}

\subsection{RQ3: Comprehensive Evaluation for Various Configurations}

In this section, we extensively analyze and evaluate the vulnerability detection performance of the combined model composed of different types of components deployed in VulFed. 
Aside from variations in input data and their preprocessing methods, the adjustable component types mainly include the selection of LLMs, FL algorithms, and PEFT strategies.
Table~\ref{peft}, Table~\ref{fl}, and Table~\ref{model} present all details of the available options for these three types of components.

VulFed encompasses a wide range of code sample structures as input.
To assess its robustness when faced with different input data structures, our main experimental part conducts tests using the same implementation process on three types of input data: raw source code, SeVC, and Code Gadget, respectively.
Specifically, we first compare the performance of each LLM deployed combined with different FL algorithms in vulnerability detection to identify superior algorithms and models.
Secondly, we compare the performance of each PEFT strategy based on specific LLM combined with different FL algorithms to determine more cost-effective fine-tuning strategies that aim to maintain low communication overhead while achieving performance close to that of full-parameter training.
Furthermore, if users choose to process raw code input into a graph structure, we independently employ the three methods described in \cite{devign}, \cite{reveal}, and \cite{vdbfl} to implement GNN-based FL vulnerability detection.
All experiments in this section are conducted on binary classification tasks.

Building upon the strategy above, we simultaneously employ horizontal and vertical comparisons for each experiment result table.
On the one hand, we aim to identify which model and corresponding PEFT method are more effective based on the same FL algorithm from the horizontal perspective.
On the other hand, vertical comparison is also important because it is crucial to discover which FL algorithm is more effective in our experimental setting when the model or fine-tuning mode is fixed.
In order to show the vertical comparison of various federated algorithms better when listing the advantages and disadvantages of the models horizontally, we annotated the table of experimental results as follows: under each LLM or PEFT strategy, the accuracy and F1 score according to the best performance among FL algorithms will be highlighted in bold, and the second highest performance among FL algorithms will be underlined.

\begin{table*}[t]
\caption{Evaluation Results of VulFed When Adopting Different LLMs Combined with FL Algorithms.}
\label{rq3-model}
 \vspace{-0.1in}
\centering
\resizebox{\textwidth}{!}{
\begin{tabular}{@{}c|c|cccccc|cccccc|cccccc@{}}
\hline\hline
\multirow{2}{*}{Alg.} & \multirow{2}{*}{Metric} & \multicolumn{6}{c|}{Perfomance of LLMs When Inputting Raw Codes} & \multicolumn{6}{c|}{Perfomance of LLMs When Inputting SeVCs} & \multicolumn{6}{c}{Perfomance of LLMs When Inputting Code Gadgets} \\ \cline{3-20} 
 &  & \multicolumn{1}{c|}{BERT} & \multicolumn{1}{c|}{CodeBERT} & \multicolumn{1}{c|}{GPT-2} & \multicolumn{1}{c|}{OPT} & \multicolumn{1}{c|}{T5} & CodeT5 & \multicolumn{1}{c|}{BERT} & \multicolumn{1}{c|}{CodeBERT} & \multicolumn{1}{c|}{GPT-2} & \multicolumn{1}{c|}{OPT} & \multicolumn{1}{c|}{T5} & CodeT5 & \multicolumn{1}{c|}{BERT} & \multicolumn{1}{c|}{CodeBERT} & \multicolumn{1}{c|}{GPT-2} & \multicolumn{1}{c|}{OPT} & \multicolumn{1}{c|}{T5} & CodeT5 \\ \hline
\multirow{4}{*}{\rotatebox{90}{FedAvg}} & Acc. & 62.01\% & \underline{63.80\%} & 60.69\% & 62.15\% & 58.20\% & 60.29\% & 74.64\% & 75.77\% & 73.91\% & 75.00\% & 68.67\% & \textbf{73.43\%} & 94.47\% & 94.86\% & 93.02\% & \underline{94.84\%} & \underline{82.99\%} & \underline{92.27\%} \\
 & Pre. & 57.60\% & 62.12\% & 57.50\% & 58.51\% & 54.26\% & 56.23\% & 50.21\% & 51.57\% & 49.40\% & 50.65\% & 43.55\% & 48.83\% & 86.08\% & 87.73\% & 81.02\% & 86.35\% & 61.70\% & 79.95\% \\
 & Rec. & 54.02\% & 54.10\% & 55.29\% & 60.56\% & 57.37\% & 61.12\% & 85.98\% & 84.65\% & 87.84\% & 83.14\% & 76.48\% & 83.85\% & 94.16\% & 93.49\% & 95.88\% & 95.45\% & 92.96\% & 94.15\% \\
 & F1 & 55.76\% & \textbf{59.75\%} & 56.38\% & 59.51\% & 55.77\% & 58.57\% & 63.40\% & 64.09\% & 63.24\% & 62.95\% & 55.06\% & \underline{61.72\%} & 89.94\% & 90.52\% & 87.83\% & \underline{90.67\%} & \textbf{74.17\%} & 86.47\% \\ \hline
\multirow{4}{*}{\rotatebox{90}{FedProx}} & Acc. & 61.17\% & 63.14\% & 58.49\% & 61.86\% & 57.24\% & 60.17\% & 73.68\% & 75.50\% & 75.02\% & 75.59\% & 67.90\% & 72.33\% & 93.69\% & 94.35\% & 92.79\% & 94.60\% & 82.96\% & 89.28\% \\
 & Pre. & 58.51\% & 61.70\% & 55.01\% & 58.34\% & 53.19\% & 56.06\% & 49.12\% & 51.27\% & 50.66\% & 51.35\% & 42.87\% & 47.16\% & 84.93\% & 87.89\% & 81.29\% & 86.78\% & 61.93\% & 72.57\% \\
 & Rec. & 53.14\% & 52.11\% & 52.91\% & 59.36\% & 57.69\% & 61.51\% & 84.29\% & 82.25\% & 85.63\% & 84.20\% & 77.20\% & 83.40\% & 92.34\% & 91.04\% & 94.25\% & 93.73\% & 92.36\% & 95.16\% \\
 & F1 & 55.70\% & 56.50\% & 53.94\% & 58.85\% & 55.35\% & \underline{58.66\%} & 62.07\% & 63.17\% & \underline{63.65\%} & 63.80\% & 55.13\% & 60.63\% & 88.48\% & 89.43\% & 87.29\% & 90.12\% & \underline{74.14\%} & 82.35\% \\ \hline
\multirow{4}{*}{\rotatebox{90}{CluSamp}} & Acc. & 61.71\% & 63.57\% & \textbf{62.52\%} & \underline{63.43\%} & \underline{58.89\%} & 61.12\% & \textbf{77.08\%} & \textbf{77.63\%} & \textbf{75.61\%} & \underline{76.36\%} & 70.67\% & 73.16\% & 94.54\% & 94.89\% & 93.34\% & 94.76\% & 81.80\% & 92.25\% \\
 & Pre. & 57.93\% & 61.73\% & 59.34\% & 61.03\% & 55.07\% & 57.52\% & 53.30\% & 53.89\% & 51.39\% & 52.29\% & 45.86\% & 48.60\% & 85.86\% & 88.01\% & 83.35\% & 87.76\% & 59.85\% & 80.11\% \\
 & Rec. & 60.79\% & 54.50\% & 58.49\% & 56.41\% & 57.14\% & 58.80\% & 83.23\% & 86.07\% & 83.58\% & 85.09\% & 81.99\% & 87.93\% & 94.49\% & 93.25\% & 93.29\% & 93.06\% & 93.29\% & 93.77\% \\
 & F1 & \textbf{59.33\%} & 57.89\% & 58.91\% & 58.63\% & \underline{56.08\%} & 58.16\% & \textbf{64.98\%} & \textbf{66.28\%} & 63.64\% & \underline{64.78\%} & 58.82\% & \textbf{62.60\%} & \underline{90.12\%} & \underline{90.56\%} & 88.05\% & 90.33\% & 72.92\% & 86.41\% \\ \hline
\multirow{4}{*}{\rotatebox{90}{FedCross}} & Acc. & 62.18\% & \textbf{64.53\%} & \underline{62.45\%} & 63.14\% & \textbf{59.18\%} & \textbf{62.08\%} & 75.61\% & 77.00\% & \underline{75.16\%} & \textbf{77.22\%} & \textbf{71.73\%} & 73.23\% & \textbf{94.81\%} & \textbf{95.03\%} & \textbf{93.47\%} & \textbf{94.98\%} & 82.20\% & 92.23\% \\
 & Pre. & 59.22\% & 63.44\% & 58.95\% & 59.84\% & 55.33\% & 58.91\% & 51.36\% & 53.09\% & 50.83\% & 53.34\% & 46.91\% & 48.62\% & 89.12\% & 88.56\% & 83.51\% & 88.26\% & 60.43\% & 79.57\% \\
 & Rec. & 56.81\% & 53.78\% & 60.07\% & 60.08\% & 57.93\% & 57.69\% & 85.36\% & 85.27\% & 84.56\% & 86.33\% & 80.92\% & 84.38\% & 91.38\% & 93.10\% & 93.63\% & 93.30\% & 93.39\% & 94.73\% \\
 & F1 & 57.99\% & 58.21\% & \underline{59.51\%} & \underline{59.96\%} & \textbf{56.60\%} & 58.29\% & \underline{64.13\%} & 65.44\% & 63.49\% & \textbf{65.94\%} & \underline{59.39\%} & 61.69\% & \textbf{90.23\%} & \textbf{90.77\%} & \textbf{88.28\%} & \textbf{90.71\%} & 73.38\% & \underline{86.50\%} \\ \hline
\multirow{4}{*}{\rotatebox{90}{Moon}} & Acc. & \textbf{62.45\%} & 63.21\% & 62.26\% & 61.97\% & 57.54\% & \underline{61.79\%} & 72.82\% & 75.63\% & 73.82\% & 74.38\% & 67.97\% & 71.55\% & 92.14\% & 92.19\% & 92.06\% & 92.41\% & 80.55\% & 91.96\% \\
 & Pre. & 59.15\% & 60.30\% & 58.31\% & 57.76\% & 53.62\% & 58.29\% & 48.19\% & 51.38\% & 49.26\% & 49.92\% & 43.10\% & 46.70\% & 80.44\% & 80.17\% & 80.37\% & 79.98\% & 58.64\% & 80.23\% \\
 & Rec. & 58.96\% & 58.33\% & 62.63\% & 64.06\% & 56.10\% & 59.12\% & 85.18\% & 85.80\% & 82.43\% & 87.76\% & 79.32\% & 80.30\% & 92.57\% & 93.34\% & 92.38\% & 94.88\% & 88.03\% & 92.10\% \\
 & F1 & \underline{59.06\%} & \underline{59.29\%} & \textbf{60.39\%} & \textbf{60.75\%} & 54.83\% & \textbf{58.70\%} & 61.56\% & 64.27\% & 61.67\% & 63.64\% & 55.86\% & 59.05\% & 86.08\% & 86.26\% & 85.94\% & 86.79\% & 70.39\% & 85.75\% \\ \hline
\multirow{4}{*}{\rotatebox{90}{FedMut}} & Acc. & \underline{62.23\%} & 63.28\% & 62.25\% & \textbf{63.80\%} & 57.93\% & 59.44\% & \underline{76.11\%} & \underline{77.52\%} & 74.05\% & 75.02\% & \underline{70.97\%} & \underline{73.30\%} & \underline{94.57\%} & \underline{94.89\%} & \underline{93.34\%} & 94.78\% & \textbf{83.56\%} & \textbf{93.08\%} \\
 & Pre. & 59.31\% & 60.66\% & 58.93\% & 60.97\% & 54.10\% & 55.69\% & 52.07\% & 53.79\% & 49.56\% & 50.63\% & 46.20\% & 48.67\% & 88.08\% & 88.61\% & 82.93\% & 87.87\% & 63.84\% & 82.56\% \\
 & Rec. & 56.57\% & 57.13\% & 58.89\% & 58.89\% & 57.93\% & 57.29\% & 81.37\% & 85.09\% & 91.04\% & 88.55\% & 83.22\% & 83.40\% & 91.71\% & 92.43\% & 94.01\% & 92.96\% & 86.25\% & 93.39\% \\
 & F1 & 57.91\% & 58.84\% & 58.91\% & 59.91\% & 55.95\% & 56.48\% & 63.50\% & \underline{65.91\%} & \textbf{64.19\%} & 64.43\% & \textbf{59.42\%} & 61.48\% & 89.86\% & 90.48\% & \underline{88.13\%} & 90.34\% & 73.38\% & \textbf{87.64\%} \\ \hline\hline
\end{tabular}

}
\end{table*}

\begin{table*}[t]
\caption{Evaluation Results of VulFed When Adopting Different PEFT Strategies Combined with FL Algorithms.}
\vspace{-0.1in}
\label{rq3-peft}
\centering
\resizebox{\textwidth}{!}{

\begin{tabular}{@{}c|c|cccccc|cccccc|cccccc@{}}
\hline\hline
\multirow{2}{*}{Alg.} & \multirow{2}{*}{Metric} & \multicolumn{6}{c|}{Perfomance of PEFT When Inputting Raw Codes} & \multicolumn{6}{c|}{Perfomance of PEFT When Inputting SeVCs} & \multicolumn{6}{c}{Perfomance of PEFT When Inputting Code Gadgets} \\ \cline{3-20} 
 &  & \multicolumn{1}{c|}{NoPEFT} & \multicolumn{1}{c|}{PTV1} & \multicolumn{1}{c|}{PTV2} & \multicolumn{1}{c|}{LoRA} & \multicolumn{1}{c|}{LoHa} & IA3 & \multicolumn{1}{c|}{NoPEFT} & \multicolumn{1}{c|}{PTV1} & \multicolumn{1}{c|}{PTV2} & \multicolumn{1}{c|}{LoRA} & \multicolumn{1}{c|}{LoHa} & IA3 & \multicolumn{1}{c|}{NoPEFT} & \multicolumn{1}{c|}{PTV1} & \multicolumn{1}{c|}{PTV2} & \multicolumn{1}{c|}{LoRA} & \multicolumn{1}{c|}{LoHa} & IA3 \\ \hline
\multirow{4}{*}{\rotatebox{90}{FedAvg}} & Acc. & \underline{63.80\%} & 54.72\% & 60.78\% & 62.23\% & \textbf{62.70\%} & \underline{57.91\%} & 75.77\% & 58.64\% & \textbf{72.55\%} & 72.57\% & \textbf{71.71\%} & 66.28\% & 94.86\% & 67.69\% & 87.26\% & \underline{92.89\%} & 89.57\% & 69.50\% \\
 & Pre. & 62.12\% & 50.68\% & 55.62\% & 59.47\% & 59.72\% & 53.99\% & 51.57\% & 35.18\% & 47.81\% & 47.86\% & 46.92\% & 43.60\% & 87.73\% & 44.32\% & 70.00\% & 82.75\% & 75.50\% & 45.59\% \\
 & Rec. & 54.10\% & 53.70\% & 56.77\% & 55.78\% & 57.76\% & 56.65\% & 84.65\% & 73.56\% & 81.46\% & 82.16\% & 81.82\% & 81.58\% & 93.49\% & 89.85\% & 90.08\% & 92.15\% & 89.27\% & 83.33\% \\
 & F1 & \textbf{59.75\%} & 52.14\% & 56.19\% & 57.56\% & \underline{58.78\%} & \textbf{55.29\%} & 64.09\% & 47.60\% & 60.26\% & 60.48\% & 59.64\% & 56.83\% & 90.52\% & 59.36\% & 78.79\% & \underline{87.20\%} & 81.80\% & 58.93\% \\ \hline
\multirow{4}{*}{\rotatebox{90}{FedProx}} & Acc. & 63.14\% & 54.46\% & 60.63\% & 60.28\% & 59.22\% & 54.17\% & 75.50\% & 58.04\% & 70.42\% & 72.05\% & \underline{71.60\%} & 65.41\% & 94.35\% & 68.07\% & \textbf{91.28\%} & 92.52\% & 88.40\% & 69.83\% \\
 & Pre. & 61.70\% & 50.41\% & 55.73\% & 57.33\% & 55.25\% & 50.13\% & 51.27\% & 34.03\% & 45.60\% & 47.32\% & 46.86\% & 42.58\% & 87.89\% & 44.57\% & 78.41\% & 82.08\% & 72.70\% & 45.78\% \\
 & Rec. & 52.11\% & 53.78\% & 54.21\% & 52.99\% & 59.12\% & 46.06\% & 82.25\% & 68.50\% & 81.99\% & 83.05\% & 83.58\% & 78.32\% & 91.04\% & 88.60\% & 92.19\% & 91.47\% & 89.42\% & 80.75\% \\
 & F1 & 56.50\% & 52.05\% & 54.96\% & 55.07\% & 57.12\% & 48.00\% & 63.17\% & 45.48\% & 58.61\% & 60.29\% & \textbf{60.05\%} & 55.18\% & 89.43\% & 59.30\% & \textbf{84.75\%} & 86.52\% & 80.19\% & 58.43\% \\ \hline
\multirow{4}{*}{\rotatebox{90}{CluSamp}} & Acc. & 63.57\% & 56.56\% & \underline{62.01\%} & 61.38\% & 59.59\% & \textbf{57.98\%} & \textbf{77.63\%} & 60.36\% & 72.34\% & \underline{72.73\%} & 70.26\% & \underline{67.68\%} & 94.89\% & \underline{68.21\%} & 90.54\% & 92.83\% & \underline{89.96\%} & 70.71\% \\
 & Pre. & 61.73\% & 50.92\% & 59.78\% & 58.70\% & 55.71\% & 54.48\% & 53.89\% & 36.25\% & 47.64\% & 47.91\% & 45.63\% & 43.13\% & 88.01\% & 44.69\% & 77.20\% & 82.83\% & 76.31\% & 46.78\% \\
 & Rec. & 54.50\% & 54.04\% & 52.83\% & 53.78\% & 58.73\% & 51.79\% & 86.07\% & 72.76\% & 83.40\% & 77.37\% & 85.80\% & 83.32\% & 93.25\% & 88.98\% & 90.80\% & 91.71\% & 89.61\% & 84.15\% \\
 & F1 & 57.89\% & 52.43\% & 56.10\% & 56.13\% & 57.18\% & 53.10\% & \textbf{66.28\%} & 48.39\% & \textbf{60.65\%} & 59.18\% & 59.58\% & \underline{56.84\%} & \underline{90.56\%} & 59.50\% & 83.45\% & 87.05\% & \underline{82.42\%} & 60.13\% \\ \hline
\multirow{4}{*}{\rotatebox{90}{FedCross}} & Acc. & \textbf{64.53\%} & \textbf{57.39\%} & \textbf{62.04\%} & \underline{62.27\%} & 60.51\% & 54.90\% & 77.00\% & \underline{60.49\%} & \underline{72.48\%} & 71.87\% & 70.13\% & 67.41\% & \textbf{95.03\%} & 68.04\% & \underline{90.71\%} & \textbf{93.14\%} & 89.75\% & 70.93\% \\
 & Pre. & 63.44\% & 51.80\% & 59.49\% & 60.04\% & 56.98\% & 50.77\% & 53.09\% & 36.66\% & 47.76\% & 47.26\% & 45.35\% & 42.66\% & 88.56\% & 44.63\% & 77.08\% & 84.53\% & 75.45\% & 47.03\% \\
 & Rec. & 53.78\% & 55.06\% & 54.42\% & 53.39\% & 57.21\% & 60.16\% & 85.27\% & 75.07\% & 82.25\% & 87.40\% & 82.61\% & 80.21\% & 93.10\% & 90.23\% & 91.95\% & 90.52\% & 90.37\% & 84.48\% \\
 & F1 & 58.21\% & \underline{53.38\%} & 56.84\% & 56.52\% & 57.10\% & \underline{55.07\%} & 65.44\% & \textbf{49.26\%} & \underline{60.43\%} & \underline{61.35\%} & 58.55\% & 55.70\% & \textbf{90.77\%} & \underline{59.72\%} & \underline{83.86\%} & \textbf{87.42\%} & 82.24\% & \underline{60.42\%} \\ \hline
\multirow{4}{*}{\rotatebox{90}{Moon}} & Acc. & 63.21\% & 56.33\% & 59.52\% & \textbf{63.80\%} & \underline{62.51\%} & 52.93\% & 75.63\% & \textbf{60.81\%} & 69.67\% & 70.85\% & 71.05\% & 66.27\% & 92.19\% & 67.71\% & 84.78\% & 87.61\% & 88.62\% & \underline{71.27\%} \\
 & Pre. & 60.30\% & 51.95\% & 55.40\% & 61.73\% & 59.26\% & 49.00\% & 51.38\% & 36.06\% & 44.72\% & 46.20\% & 46.34\% & 42.13\% & 80.17\% & 43.99\% & 65.27\% & 70.34\% & 72.36\% & 47.33\% \\
 & Rec. & 58.33\% & 65.90\% & 60.88\% & 55.78\% & 58.88\% & 60.88\% & 85.80\% & 69.12\% & 79.33\% & 85.99\% & 84.38\% & 85.80\% & 93.34\% & 84.01\% & 89.85\% & 91.33\% & 91.67\% & 83.19\% \\
 & F1 & \underline{59.29\%} & \textbf{58.09\%} & \textbf{58.01\%} & \textbf{58.60\%} & \textbf{59.07\%} & 54.30\% & 64.27\% & 47.41\% & 57.20\% & 60.11\% & \underline{59.83\%} & 56.52\% & 86.26\% & 57.74\% & 75.61\% & 79.47\% & 80.88\% & 60.33\% \\ \hline
\multirow{4}{*}{\rotatebox{90}{FedMut}} & Acc. & 63.28\% & \underline{56.84\%} & 60.00\% & 60.72\% & 59.98\% & 54.24\% & \underline{77.52\%} & 59.95\% & 71.91\% & \textbf{73.59\%} & 69.67\% & \textbf{70.01\%} & \underline{94.89\%} & \textbf{70.23\%} & 88.30\% & 92.04\% & \textbf{90.29\%} & \textbf{78.77\%} \\
 & Pre. & 60.66\% & 53.00\% & 56.02\% & 56.83\% & 55.96\% & 50.18\% & 53.79\% & 36.30\% & 47.10\% & 49.01\% & 44.69\% & 45.21\% & 88.61\% & 46.28\% & 72.16\% & 80.45\% & 76.36\% & 57.23\% \\
 & Rec. & 57.13\% & 53.47\% & 60.08\% & 60.32\% & 60.56\% & 55.86\% & 85.09\% & 75.24\% & 80.83\% & 83.05\% & 78.70\% & 82.08\% & 92.43\% & 83.00\% & 90.28\% & 92.05\% & 91.28\% & 75.81\% \\
 & F1 & 58.84\% & 53.23\% & \underline{57.98\%} & \underline{58.52\%} & 58.17\% & 52.87\% & \underline{65.91\%} & \underline{48.97\%} & 59.52\% & \textbf{61.64\%} & 57.01\% & \textbf{58.30\%} & 90.48\% & \textbf{59.42\%} & 80.21\% & 85.86\% & \textbf{83.16\%} & \textbf{65.22\%} \\ \hline\hline
\end{tabular}

}
\end{table*}

{\bf Impacts of Different LLMs.}
Table~\ref{rq3-model} shows the evaluation results of deploying different LLMs combined with FL algorithms. 
By analyzing this result, we can directly draw conclusions on how to wisely choose LLM, which is generally applicable to most FL algorithms and various types of input data.
In decoder-only LLMs, CodeBERT, due to its improved pertaining data and process design for code-related tasks, consistently outperforms BERT in vulnerability detection tasks when using versions of similar scale, achieving approximately a 1\% lead in both accuracy and F1 Score.
Among encoder-only LLMs, GPT-2 and OPT employ the lightweight, small version.
However, OPT, a series of models designed with GPT-3 as its target released closer to the present, utilized a wider corpus and better strategies during pre-training, thus consistently outperforming GPT-2 in almost all scenarios. 
In encoder-decoder structure LLMs, CodeT5 significantly outperforms T5, although it generally lags behind other categories of LLMs.
When using the most lightweight small version, the parameter count of the T5 series is only half that of other LLMs, which limits the upper bound of its performance to some extent.
Their performance can be improved if the lightweight restriction is disregarded and the base version is adopted.

{\bf Impacts of Different PEFT Strategies.}
Table~\ref{rq3-peft} shows the evaluation results of adopting different PEFT strategies combined with FL algorithms.
In the same way as when analyzing LLM, we mainly consider the performance advantages and disadvantages of different types of PEFT strategies when confronting various combinations of other components.
To control unnecessary variables, all LLMs trained with various PEFT strategies in the figure are CodeBERT, so the data in the ``NoPEFT'' column comes from the corresponding data in Table~\ref{rq3-model}.
In most cases, the PEFT strategies P-Tuning V2 and LoRA achieve performance closest to that of full-parameter fine-tuning, with the former generating greater communication overhead than the latter due to adding an MLP layer.
P-Tuning V1, as an older version of P-Tuning V2, generally struggles to achieve comparable performance. IA3 shows a similar pattern compared to LoRA, exhibiting better performance in a few cases.
At the same time, LoHa, another PEFT strategy akin to LoRA, generally performs similarly to LoRA in most cases and only surpasses LoRA in rare cases.

{\bf Impacts of Different FL Algorithms.}
The performance of various FL algorithms is closely linked to the selection of other components. Table~\ref{rq3-model} and Table~\ref{rq3-peft} present the best-performing and second-best-performing algorithms under different component configurations. 
FedAvg, the first FL algorithm proposed, is used as a baseline for comparing with other methods. In VulFed, FedAvg's overall performance is relatively low compared to all algorithms. 
FedProx, an improved version of FedAvg, generally performs poorly and even fails to match FedAvg's standard. 
Compared to the former two methods, Cluster Sampling exhibits superior capability. It achieves top performance in detecting SeVC vulnerabilities regardless of the deployed LLMs and maintains leading performance in other scenarios. 
FedCross is the most universal and performs best among all algorithms. It demonstrates a significant advantage in detecting Code Gadget and source code vulnerabilities, although its convergence speed is generally slower than other algorithms. 
The overall performance of Moon is on average, but does not outperform FedAvg in some cases. However, its performance is particularly remarkable when employing PEFT strategies to detect vulnerabilities in raw code. 
FedMut is also an algorithm with outstanding performance, consistently maintaining good performance across all scenarios. It holds a significant advantage in detecting code gadget vulnerabilities, particularly when using LoHa and IA3. 
Overall, FedCross, FedMut, and Cluster Sampling are superior in comprehensive performance.

\begin{table}[ht]
\caption{Evaluation Results of GNN-based Methods.}
\vspace{-0.05in}
\label{gnn}
\centering
\tiny
\begin{tabular}{c|c|ccc|c|c|ccc}
\hline\hline
\multirow{2}{*}{FL Algorithm} & \multirow{2}{*}{Metrics} & \multicolumn{3}{c|}{Performance} & \multirow{2}{*}{FL Algorithm} & \multirow{2}{*}{Metrics} & \multicolumn{3}{c}{Performance} \\ \cline{3-5} \cline{8-10} 
 &  & Devign & Reveal & VDBFL &  &  & Devign & Reveal & VDBFL \\ \hline
\multirow{4}{*}{FedAvg} & Accuracy & 63.10\% & 70.79\% & 74.08\% & \multirow{4}{*}{FedCross} & Accuracy & \textbf{64.13\%} & \textbf{71.74\%} & 74.08\% \\
 & Precision & 57.95\% & 68.24\% & 71.65\% &  & Precision & 58.93\% & 68.91\% & 70.19\% \\
 & Recall & 71.71\% & 68.12\% & 72.11\% &  & Recall & 72.27\% & 70.12\% & 76.57\% \\
 & F1 Score & 64.10\% & 68.19\% & 71.88\% &  & F1 Score & \textbf{64.92\%} & \textbf{69.51\%} & 73.24\% \\ \hline
\multirow{4}{*}{FedProx} & Accuracy & 62.51\% & 69.14\% & 73.57\% & \multirow{4}{*}{Moon} & Accuracy & 62.88\% & 70.93\% & 73.79\% \\
 & Precision & 57.43\% & 65.58\% & 70.30\% &  & Precision & 57.76\% & 67.42\% & 69.18\% \\
 & Recall & 71.16\% & 69.08\% & 73.55\% &  & Recall & 71.39\% & 71.08\% & 77.45\% \\
 & F1 Score & 63.56\% & 67.29\% & 71.88\% &  & F1 Score & 63.87\% & 69.20\% & 73.08\% \\ \hline
\multirow{4}{*}{CluSamp} & Accuracy & 63.43\% & 70.49\% & 74.30\% & \multirow{4}{*}{FedMut} & Accuracy & 64.09\% & 71.70\% & \textbf{74.93\%} \\
 & Precision & 58.04\% & 67.91\% & 70.19\% &  & Precision & 59.23\% & 69.53\% & 70.83\% \\
 & Recall & 73.63\% & 67.81\% & 76.57\% &  & Recall & 70.04\% & 68.36\% & 77.21\% \\
 & F1 Score & 64.91\% & 67.86\% & 73.24\% &  & F1 Score & 57.91\% & 68.94\% & \textbf{73.88\%} \\ \hline\hline
\end{tabular}
\vspace{-0.15pt}
\end{table}

{\bf Evaluation Results of GNN-based Models.}
VulFed is capable of accepting graph-structured code samples as inputs.
However, due to the utilization of GNN-based models instead of LLMs, this component configuration has lower relevance to other strategies.
Table~\ref{gnn} shows the performance of three types of GNN-based methods combined with FL algorithms. 
On the one hand, since Reveal improves based on Devign and VDBFL further adapts these methods to FL scenes, their performance rises in a gradual manner.
On the other hand, FL algorithms show a comprehensive performance differentiation similar to that when utilizing LLM when processing this task.
FedCross and FedMut are the two best methods for performance.

\begin{tcolorbox}
    [
    colback=yellow!7!white,
    colframe=white!75!white,
    width=\textwidth,
    arc=1mm, auto outer arc,
    boxrule=0.5pt,
    ]
{\bfseries Answer to RQ3:}
Comprehensive evaluation results indicate that CodeBERT and OPT lead in all deployable models, while P-Tuning V2 and LoRA, as PEFT Strategies, can train a model whose performance is closest to full parameter fine-tuning. Although FL algorithms perform diversely in different component configurations, Fedcross, FedMut, and Cluster Sampling are the most recommended methods according to their comprehensive performance.
\end{tcolorbox}

\section{Discussion}
\label{sec:discussion}

\subsection{Extension of Framework Components}
When designing VulFed, we priorly choose representative methods or models as optional framework components.
Firstly, in terms of the model we deployed, we comprehensively considered the requirements of the FL framework for lightweight models and the necessity of LLM maintaining its type and then selected six most typical LLMs under three different architectures.
At present, the performance and parameter scale of LLM are developing rapidly, but we have not adopted the latest and most parameterized models, such as Llama2.
On one hand, excessive model parameters can significantly increase computational overhead, and combined with the computational requirements generated by testing a newly deployed model in our study, it can increase the time cost to an unacceptable standard.
On the other hand, the typical models we have chosen are highly representative. When these models perform well in the evaluation phase, it proves the effectiveness of our framework and the correctness of our research conclusions, which can be further extended to better models.
In addition, we adhere to the same philosophy when selecting the PEFT method and FL algorithm.
Among all the parameter-efficient fine-tuning methods, prompt tuning and parameter low-rank decomposition are the most representative.
We have selected typical methods for each stage and deployed them to the framework.
Similarly, we select typical methods from various FL optimization algorithms, including clustering-based, cross-aggregation-based, and contrastive-learning-based.
We believe there is no need to test too many methods in the evaluation process.
However, it should be noted that all components in VulFed are plug-and-play, and any new method can be quickly added to the optional range of our framework after certain processing, which benefits VulFed to follow up on new achievements in future related research rapidly.

\subsection{Problems of Multilingual Sources of Vulnerability Samples}
In this section, we discuss three potential directions for further study of VulFed.
Firstly, since the components in VulFed are organized in a pluggable manner, it can be easily extended to include a wider range of deployable models, selectable PEFT strategies, and FL algorithms.
Secondly, while all evaluation tasks currently accept code samples from the C/C++ programming languages in the dataset, cross-language tasks may become a new requirement in the joint scenario proposed in this paper: suppose the privacy vulnerability codes provided by different clients are written in different programming languages, then the global model in FL needs to have the capability for cross-language vulnerability detection. 
Therefore, the code must be transformed into a unified intermediate form at the preprocessing stage before training the locally deployed model. However, whether such a strategy can effectively enhance the overall detection capability of the model remains to be studied.

\subsection{Problems of Fine-grained Classification}
Our study utilizes two classification tasks: a binary classification for detecting whether the target code sample has vulnerabilities and a multi-classification for detecting which type of vulnerabilities the target code sample possesses.
Considering the significant efforts and computation costs we have already incurred, we did not further expand the compatibility of VulFed in classification tasks.
However, current vulnerability detection tasks already have more fine-grained classification strategies, such as detection that is precise to the line of code or a certain variable.
In real-world scenarios, more fine-grained vulnerability detection methods are often more in demand.
For example, when detecting code samples with long text lengths, it is necessary to obtain more accurate information on the location of the potential vulnerability.
In addition, if the model directly generates a smaller potential vulnerability area, it can help to improve task efficiency and reduce labor costs.
However, it remains to be verified whether a detection task with more fine granularity can achieve better results under the FL framework.
Such tasks are usually accompanied by more complex preprocessing and training steps, which must be well integrated with FL to generate benefits.
These issues are all worth further consideration.
In the future, we plan to conduct a more in-depth analysis and evaluation of the possibility of combining more fine-grained vulnerability detection with the FL system and further expand our framework within feasible limits to enable it to handle different vulnerability detection tasks to meet more needs.


\section{Conclusion}
\label{sec:conclusion}

Along with the prosperity of AI techniques, Federated Learning (FL) is becoming a promising collaborative learning scheme for software vulnerability detection since it can effectively address the data silo problem among relevant companies or departments without compromising their data (software) privacy. 
However, existing FL-based vulnerability detection methods focus on application-specific tasks with GNN-like inputs, limiting their usage in common vulnerability detection tasks. 
To reveal the potential of FL in this domain, in this paper, we proposed a novel FL vulnerability detection framework named VulFed that supports both traditional DL models and rapidly evolving LLMs.
Based on VulFed, we revealed that FL-based methods can significantly improve overall vulnerability detection performance at the CWE-case level for both IID and non-IID scenarios.
Meanwhile, we evaluated the performance of different data processing and FL optimization methods (e.g., data preprocessing mechanisms, FL training schemes,  PEFT) on VulFed.
Comprehensive experimental results obtained from the study reveal the impacts of such different FL configurations on various vulnerability detection tasks, which can help researchers build their own FL frameworks to address specific vulnerability detection tasks. 
In the future, we plan to investigate larger models and their performance optimization mechanisms based on our VulFed framework. Moreover, the finer-grained (e.g., line or character level) and cross-language vulnerability detection tasks are also worth further study.



\newpage

\bibliography{reference}

\end{document}